\documentclass[11pt]{article}
\usepackage{axodraw}
\usepackage{epsfig}
\usepackage{amsfonts}
\usepackage{amsmath}
\usepackage{bbm}
\input pix.sty

\renewcommand{\eq}{eq.~}
\renewcommand{\eqs}{eqs.~}
\renewcommand{\se}{sec.~}

\renewcommand{\fig}{fig.~}

\newcommand{\tinymsbar}{{\overline{\mbox{\tiny\rm{MS}}}}}
\newcommand{\Lambdamsbar}{{\Lambda_\tinymsbar}}

\newcommand{\Nf}{N_{\rm f}}
\newcommand{\Nc}{N_{\rm c}}

\newcommand{\mE}{m_\rmii{E}}
\newcommand{\gE}{g_\rmii{E}}
\newcommand{\gammaE}{\gamma_\rmii{E}}

\newcommand{\rmO}{{\mathcal{O}}}
\newcommand{\bmu}{\bar\mu}

\def\lsi{\raise0.3ex\hbox{$<$\kern-0.75em\raise-1.1ex\hbox{$\sim$}}}
\def\gsi{\raise0.3ex\hbox{$>$\kern-0.75em\raise-1.1ex\hbox{$\sim$}}}

\newcommand{\rmii}[1]{{\mbox{\tiny\rm{#1}}}}
\newcommand{\re}{\mathop{\mbox{Re}}}
\newcommand{\im}{\mathop{\mbox{Im}}}

\newcommand{\Tint}[1]{{\hbox{$\sum$}\!\!\!\!\!\!\!\int\,}_{\!\!\!\!\raise-0.9ex\hbox{$\scriptstyle{#1}$}}}
\newcommand{\Tinti}[1]{{{\Sigma}\!\!\!\!\raise0.3ex\hbox{$\int$}_\rmii{${#1}$}}}

\newcommand{\bi}{\begin{itemize}}
\newcommand{\ei}{\end{itemize}}

\newcommand{\hide}[1]{ }

\def\TAsc(#1,#2)(#3,#4,#5)%
{\SetWidth{2.0}\CArc(#1,#2)(#3,#4,#5)\SetWidth{1.0}}
\def\Lwidth{3}

\def\TAgl(#1,#2)(#3,#4,#5){\SetWidth{2.0}\PhotonArc(#1,#2)(#3,#4,#5){\Lwidth}%
{6.283 #3 mul 360 div #4 #5 sub #4 #5 sub mul sqrt mul Tdensity mul}%
\SetWidth{1.0}}
\def\TLgl(#1,#2)(#3,#4){\SetWidth{2.0}\Photon(#1,#2)(#3,#4){\Lwidth}
{#1 #3 sub #1 #3 sub mul #2 #4 sub #2 #4 sub mul add sqrt Tdensity mul}%
\SetWidth{1.0}}
\newcommand{\piC}[1]{\;\parbox[c]{40pt}{\begin{picture}(120,60)(0,-20)
\SetWidth{1.0}\SetScale{0.35} #1 \end{picture}}\;}
\def\ConnectedA(#1,#2,#3){\piC{#1(60,-15)(75,34,146) #2(60,75)(75,214,326)%
 #3(60,60)(20,190,350)%
 \GBoxc(0,30)(10,10){1} \GBoxc(120,30)(10,10){1}%
  }}
\def\ConnectedB(#1,#2,#3){\piC{#1(60,-15)(75,34,146) #2(60,75)(75,214,326)%
 #3(60,60)(60,0)%
 \GBoxc(0,30)(10,10){1} \GBoxc(120,30)(10,10){1}%
  }}
\def\ConnectedC(#1,#2){\piC{#1(60,-15)(75,34,146) #2(60,75)(75,214,326)%
 \GBoxc(0,30)(10,10){1} \GBoxc(120,30)(10,10){1}%
  }}
\def\ConnectedD(#1,#2){\piC{#1(60,-15)(75,34,146) #2(60,75)(75,214,326)%
 \GBoxc(0,30)(10,10){1} \GBoxc(120,30)(10,10){1}%
 \SetWidth{2.0} 
 \Line(55,55)(65,65)%
 \Line(55,65)(65,55)
  }}

\makeatletter
\renewcommand\section{\@startsection {section}{1}{\z@}%
                                   {-5.5ex \@plus -1ex \@minus -.2ex}
                                   {2.3ex \@plus.2ex}%
                                   {\normalfont\large\bfseries}}
\renewcommand\subsection{\@startsection{subsection}{2}{\z@}%
                                     {-3.25ex\@plus -1ex \@minus -.2ex}%
                                     {1.5ex \@plus .2ex}%
                                     {\normalfont\normalsize\bfseries}}
\renewcommand\thesection {\@arabic\c@section}
\renewcommand\thesubsection   {\thesection.\@arabic\c@subsection}
\renewcommand{\@seccntformat}[1]{%
\csname the#1\endcsname.\hspace{1.0em}}
\makeatother

\begin{document}

\begin{titlepage}

\begin{flushright}
\vspace*{1cm}
\end{flushright}
\begin{centering}
\vfill

{\Large{\bf
On the smallest screening masses in hot QCD 
}} 

\vspace{0.8cm}

M.~Laine$^{\rm a}$, 
M.~Veps\"al\"ainen$^{\rm b}$ 

\vspace{0.8cm}

$^\rmi{a}$%
{\em
Faculty of Physics, University of Bielefeld, 
D-33501 Bielefeld, Germany\\}

\vspace{0.3cm}

$^{\rm b}$%
{\em 
Department of Physics, 
P.O.Box 64, FI-00014 University of Helsinki, Finland\\}

\vspace*{0.8cm}

\mbox{\bf Abstract}
 
\end{centering}

\vspace*{0.3cm}
 
\noindent
The increasing focus on unquenched lattice simulations 
has revived interest also in gluonic screening masses, whose 
inverses characterise the longest length scales at which thermal
fluctuations are correlated in a hot non-Abelian plasma. We fill an apparent 
gap in the literature concerning the theoretical structure of one of the 
relevant screening masses (the one which equals twice the Debye mass
at leading order), by showing that the next-to-leading order 
correction to it is perturbative and small. This surprising result
appears to explain semi-quantitatively why this particular channel yields 
the smallest gluonic screening mass at temperatures around 
a few hundred MeV (it couples to the energy density 
and to the real part of the Polyakov loop), 
even though it is not among the smallest
screening masses at asymptotically high temperatures. 

\vfill

 
\vspace*{1cm}
  
\noindent
September 2009

\vfill

\end{titlepage}

%
\section{Introduction}

Screening masses, or inverses of equal-time correlation lengths, 
are a fundamental characteristic of the long-range properties of
a thermal system. Indeed, the quantum numbers and the degeneracy of
the excitation with the lowest screening mass indicate what kind of 
an effective theory it is that determines the infrared sensitive 
thermodynamic properties of the system,
such as finite volume effects~\cite{hm1}. 
In QED, for instance,
correlators of magnetic fields display a vanishing screening mass, while 
correlators of electric fields reveal a non-vanishing ``Debye mass'';
this then shows that at the longest length scales only magnetic
fields are significant in an Abelian plasma, and finite-volume
effects are powerlike.   

In non-Abelian gauge theories such as QCD, it turns out that the 
situation with the screening masses is a bit more complicated
than in QED. In fact, even the definition of what is meant by 
screening masses requires some care: electric and magnetic fields, 
on which our Abelian intuition is based, are no longer gauge-invariant
objects. Because of these subtleties it was only in the mid-1990's
that fully satisfactory  
gauge-invariant and non-perturbative definitions were
given to colour-electric, colour-magnetic, and certain more refined
classes of screening masses in a non-Abelian plasma~\cite{ay}.

Following the conceptual clarification of the gauge-invariant
definition of gluonic
screening masses in QCD, systematic lattice measurements
could also be carried out in all relevant channels. We would like
to mention, in particular, 
quenched lattice measurements in four dimensions~\cite{4d,hm2}; 
unquenched lattice measurements via a dimensionally reduced effective 
field theory in three dimensions~\cite{mu}; 
and, most recently, unquenched lattice measurements 
directly in four dimensions~\cite{eh}. Of course, 
a systematic analysis of the same observables can also  
be carried out in the AdS/CFT framework~\cite{bky}.

The purpose of the present paper is to consider the screening 
masses within the weak-coupling expansion. A number of them fall 
into the general class of observables whose leading-order value is 
fixed by the Debye scale; this class includes also many real-time
observables of current interest, such as heavy-quark diffusion and 
jet quenching. It has been found in several such cases that the 
next-to-leading order correction is large for phenomenologically
interesting values of the gauge coupling~\cite{mink}. Our results
will produce a ``counter-example'' to this empirical
observation, showing that it is also possible to find 
observables in which the next-to-leading order correction is small.


%
\section{General framework}

In order to implement the resummations that are needed for 
defining the weak-coupling expansion at high temperatures, we choose to view
the screening masses with the help 
of the dimensionally reduced effective field theory for hot QCD~\cite{dr}, 
called EQCD~\cite{bn}. 
This approach is certainly sufficient for clarifying the 
theoretical structure of the various screening masses
and, at least on the semi-quantitative level, 
also for numerical estimates. 
The effective Lagrangian has the form 
\be
 {\mathcal{L}}_\rmii{E}  =  
 \fr12 \tr [F_{ij}^2 ]+ \tr [D_i,A_0]^2 + 
 \mE^2\tr [A_0^2] 
 + ... 
 \; . 
 \hspace*{0.5cm} \la{eqcd}
\ee
Here 
$F_{ij} = (i/\gE) [D_i,D_j]$, 
$D_i = \partial_i - i \gE A_i$, 
$A_i = A^a_i T^a$, 
$A_0 = A^a_0 T^a$, 
and $T^a$ are hermitean generators of SU(3). 
In three dimensions the dimensionality of $\gE^2$ is GeV.
A 2-loop derivation of $\mE^2$, $\gE^2$ in terms 
of the parameters of four-dimensional QCD can be found in ref.~\cite{gE2}.

Correlation lengths are defined from the exponential fall-off of 
two-point functions of local gauge-invariant operators. Without a loss
of generality we assume the two-point functions to be measured in 
the $x_3$-direction. The independent channels can be classified 
according to discrete symmetries defined in the two-dimensional
transverse $(x_1,x_2)-$plane. A particularly important symmetry is often 
called {R}, and corresponds  
to the {CT}-symmetry of the original 
QCD; in terms of \eq\nr{eqcd}, it sets $A_0\to -A_0$. ``Colour-electric
operators'' are defined to be odd under this symmetry, while 
``colour-magnetic operators'' are even~\cite{ay}.

With this notion, examples of operators from which 
colour-electric screening masses can be determined 
are $\tr[A_0 F_{12}]$ and $\tr[A_0^3]$.
In four-dimensional QCD, these correspond to 
$\im \tr[P F_{12}]$ and $\im\tr[P]$, respectively, 
where $P$ is the yet untraced Polyakov loop (we assume that 
the center symmetry is broken in the ``trivial'' direction, 
as is certainly the case in the unquenched theory). 
Note that these two 
channels do not couple to each other because of a different
parity in the transverse plane. 
Typical operators from which colour-magnetic screening
masses can be determined are $\tr[A_0^2]$ and $\tr[F_{12}^2]$, 
but any other gauge-invariant local singlet operator such as 
the energy density works as well. 
In four-dimensional QCD, $\tr[A_0^2]$ corresponds to $\re\tr[P]$.

It is important to note that in principle the operators 
$\tr[A_0^2]$ and $\tr[F_{12}^2]$ couple 
to each other~\cite{sn2,bn2}. In other
words, if we measure a correlation matrix between these operators, 
then the matrix includes non-diagonal components. It is possible, 
however, to diagonalize the correlation matrix at large distances, i.e.\
to find two orthogonal eigenstates which display different
screening masses (see, e.g., refs.~\cite{matrix}). 
It is these eigenvalues of the two-dimensional 
Hamiltonian that we refer to as $M_2$ and $M_3$ in the following. In practice, 
the coupling between $\tr[A_0^2]$ and $\tr[F_{12}^2]$
is very weak, both parametrically~\cite{bn2} and 
numerically~\cite{mu}, so it appears to us that 
it should play no actual role in our analysis. 

Assuming that $\mE  \gg \gE^2/\pi$, as is  
indeed the case at very high temperatures (in which limit
$\mE\approx gT(\Nc/3+\Nf/6)$, $\gE^2\approx g^2 T$, 
where $g^2/4\pi = \alpha_s$ is the strong 
gauge coupling, $\Nc$ is the number of colours, 
and $\Nf$ is the number of massless quark flavours), 
we can view $A_0$ as a heavy field and write down
the parametric forms of various screening masses \pagebreak
within the heavy-mass expansion\footnote{%
 We stress that at this point 
 the scale hierarchy $\mE  \gg \gE^2/\pi$ serves only as 
 a theoretical organizing principle for the computation; 
 in practical estimates various group theory and numerical factors  
 need to be added, and the phenomenological viability of the  
 description can only be estimated {\em a posteriori}.
  }. 
In particular, the smallest screening mass in the colour-electric
channel, coupling to $\tr[A_0 F_{12}]$, has a well-known logarithmic
term at the next-to-leading order~\cite{ar}, 
and the general form~\cite{ay}
\be
 M_1 \approx \mE + \frac{\gE^2\Nc}{4\pi}
 \biggl( \ln\frac{\mE}{\gE^2}  +c_1 \biggr)
 \;, \la{M1_def}
\ee
where $c_1 \approx 6.9$ for $\Nc = 3$~\cite{lp2}. 
This expression works reasonably
well down to low temperatures, overestimating the ``exact'' value within EQCD
by a modest amount~\cite{mu}. In the colour-magnetic channel, 
we can expect the mass coupling to $\tr[A_0^2]$ 
to have, in the heavy-mass limit, the form
\be
 M_2 \approx 2 \mE + \frac{\gE^2\Nc}{4\pi}
 \biggl( \ln\frac{\mE}{\gE^2}  +c_2 \biggr)
 \;. \la{M2_def} 
\ee
Roughly, 
the correction here represents a three-dimensional bosonic analogue 
of the binding energy of a heavy quark-antiquark system, like $J/\psi$.
As far as we can see it is a non-trivial fact, following from the analysis 
in \se\ref{se:M2}, that the coefficient of the logarithm 
in \eq\nr{M2_def} agrees with that in \eq\nr{M1_def}.  
The colour-magnetic screening mass which is the smallest at asymptotically 
high temperatures can, in contrast,  
be obtained from the theory from which $A_0$ 
has been integrated out~\cite{bn2};
it couples dominantly to $\tr[F_{12}^2]$  
and has the form
\be
 M_3 \approx \frac{\gE^2\Nc}{4\pi} \times c_3
 \;, \la{M3_def}
\ee
where $c_3 \approx 10.0$ for $\Nc = 3$~\cite{mt}.


%
\section{Non-relativistic limit}

Our goal now is to estimate the coefficient $c_2$ in \eq\nr{M2_def} which, 
to the best of our knowledge, remains unknown. 
The situation is quite similar to that 
in the case of the screening masses of fermionic bilinears, which 
we have studied previously in refs.~\cite{lv,mv}. The two adjoint
scalar fields form a bound state, and a formal scale hierarchy exists 
between the heavy scalar mass, $\mE$; the relative momentum between
the bound state constituents, 
$p \sim (\gE^2 \mE/\pi)^\fr12$; and the binding 
energy, $\Delta E \sim \gE^2/\pi$, such that $p^2/\mE\sim \Delta E$ 
(logarithms and numerical factors have been omitted; 
note that in terms of the four-dimensional coupling 
the scales are separated only by $\sim (g/\pi)^{1/2}$). 
This scale hierarchy can be employed for constructing a set of effective
field theories, perhaps ultimately a scalar analogue of PNRQCD~\cite{pnrqcd}. 
As argued in ref.~\cite{lv}, however, at the level of the correction 
of $\rmO(\gE^2/\pi)$, 
the whole procedure simply amounts to solving the 
Schr\"odinger equation in a two-dimensional 
Coulombic potential for the s-wave state; 
the only complication is that the heavy constituent ``rest mass'' 
entering the bound state problem needs to be fixed by a proper
matching computation. 

To nevertheless 
give a somewhat more concrete indication of the effective theory
setup, let us carry out a Wick rotation from the 3-dimensional 
Euclidean theory to a (2+1)-dimensional Minkowskian theory, and rename 
the $x_3$-coordinate to be time, $t$. Let us, furthermore, write the time
dependence of the 
quadratic part of the action in Fourier space, with $\omega$ denoting
the frequency:
\be
 S_\rmii{E} = \int_{\omega}\int_{\vec{x}} \tr\Bigl\{ A_0(-\omega,\vec{x})
 \Bigl[-\omega^2  + \mE^2 - \nabla^2
 \Bigr] A_0(\omega,\vec{x}) \Bigr\}
 + \ldots 
 \;. 
\ee 
If we concentrate on modes close to the on-shell points, 
$|\omega \pm \mE| \sim \gE^2/\pi \ll \mE$, 
and write $\omega = \mE + \omega'$
or $\omega = -\mE - \omega'$, 
then we observe that the dynamics of the forward-propagating mode
$A_0'(\omega',\vec{x}) \equiv A_0(\mE + \omega',\vec{x})$
and the backward-propagating mode 
$A_0'^\dagger(\omega',\vec{x}) = A_0(-\mE - \omega',\vec{x})$ 
is determined by the non-relativistic Lagrangian
\be
 \mathcal{L}_\rmii{E} \approx 2 \mE \tr \Bigl\{ 
 A_0'^\dagger 
 \Bigl(-i \partial_t - \frac{\nabla^2}{2 \mE} \Bigr) A_0' 
 + 
 A_0' 
 \Bigl(i \partial_t - \frac{\nabla^2}{2 \mE} \Bigr) A_0'^\dagger 
 \Bigr\}
 \;. \la{hset}
\ee
In configuration space, the original field $A_0$ is related 
to the new effective fields by 
$
 A_0 = e^{-i \mE t} A_0' +  e^{i \mE t} A_0'^\dagger
$. 
At leading order, then, the forward-propagating part of the composite 
operator $\tr[A_0^2(t)]$ has the energy eigenvalue $2 \mE$.

When this argumentation is promoted to the quantum level, we expect 
the derivatives appearing in \eq\nr{hset} to get replaced by covariant 
derivatives, 
$\partial_t A_0' \to [D_t, A_0']$; 
the rapid oscillation frequency $\mE$ to get replaced by
a matching coefficient, which we denote by $M_\rmi{rest}$; and the
parameter $\mE$ in the denominator of the kinetic term in \eq\nr{hset} 
to become another matching coefficient, which we 
denote by $M_\rmi{kin}$. In the limit $M_\rmi{kin} \to \infty$
the propagators of the $A_0'$'s are replaced by Wilson lines 
in the adjoint representation: 
$
 G(t,\vec{r}) \equiv
 \left\langle 
   A_0'^a(t,\vec{r}) A_0'^a(t,\vec{0}) 
   A_0'^b(0,\vec{r}) A_0'^b(0,\vec{0}) 
 \right\rangle 
 = \tr \{ U_\rmi{adj}(t,\vec{r}) U^T_\rmi{adj}(t,\vec{0}) \}
$, 
where $U_\rmi{adj}(t,\vec{r})$ is a straight timelike adjoint Wilson line
at spatial position $\vec{r}$ and we have for brevity 
omitted the (non-unique) spacelike connectors that 
make the point-split 
operators gauge-invariant. The evaluation of this 
expectation value leads to the concept of a static 
potential in the usual way: 
 $
  V(\vec{r}) = \lim_{t\to\infty} 
 [i\partial_t G(t,\vec{r})] G^{-1}(t,\vec{r}) 
 $. 
For the actual bound state 
problem $M_\rmi{kin} \approx \mE$ stays finite and 
the static potential takes the role of a matching coefficient. 
We do not need to know more about the effective theory setup 
in the following but remark that a formal discussion
can be found in ref.~\cite{hset}.


%
\section{Determination of $M_2$}
\la{se:M2}

Proceeding now with the non-relativistic setup outlined above, 
we expect that in the heavy mass limit
the bound state mass can be written as 
\be
 M_2 \approx 2 M_\rmi{rest} + \Delta E 
 \;. \la{DeltaE}
\ee
In dimensional regularization in $d=3-2\epsilon$ spatial 
dimensions, the next-to-leading order value 
of the matching coefficient $M_\rmi{rest}$ reads~\cite{ay} 
\be
 M_\rmi{rest} = \mE - \frac{\gE^2\Nc}{8\pi}
 \biggl( \frac{1}{\epsilon} 
 + \ln \frac{\bmu^2}{4\mE^2} + 1
 \biggr) 
 \;, \la{matching}
\ee
where $\bmu$ is the scale parameter of the $\msbar$ scheme. 
The binding energy can be solved from a two-dimensional 
Schr\"odinger equation; the potential appearing in it, obtained by 
integrating out the time (or $x_3$) components of the gauge fields reads
\be
 V(\vec{r}) = \gE^2\Nc 
 \int \! \frac{{\rm d}^{2-2\epsilon}\vec{q}}{(2\pi)^{2-2\epsilon}}
 \, \frac{1-e^{i \vec{q}\cdot\vec{r}} }{\vec{q}^2}
 = \frac{\gE^2\Nc}{4\pi}
 \biggl(\frac{1}{\epsilon} + \ln\frac{\bmu^2 r^2}{4} + 
 2 \gammaE \biggr) 
 \;. \la{Vr}
\ee
In total, then, we are looking for the ground state solution to the problem 
\be
 \biggl[
    2 M_\rmi{rest} - \frac{\nabla_\vec{r}^2}{\mE} + V(\vec{r}) 
 \biggr] \Psi_0
 = M_2 \Psi_0
 \;, \la{Seq} 
\ee
where the non-kinetic terms combine to the finite expression
\be
 2 M_\rmi{rest} + V(\vec{r}) = 
 2 \mE + \frac{\gE^2\Nc}{2\pi}
 \biggl[
    \ln(\mE r) + \gammaE - \fr12  
 \biggr]
 \;. \la{fullVr}
\ee
In the kinetic term of \eq\nr{Seq}, we already expanded the 
(``reduced'' version of the) matching 
coefficient $M_\rmi{kin}$ to leading order in 
$\gE^2/\pi\mE$, as is sufficient at 
the current level of accuracy. 

It is important to note that, unlike speculated in earlier
works~\cite{irdiv}, no infrared divergences appear in \eq\nr{fullVr}.
The reason is that  
the logarithmic divergences originating from 
the ``hard'' momenta ($q\sim \mE$; 
\eq\nr{matching}; viewed from this side $1/\epsilon$ is an infrared 
divergence)
and the ``soft'' momenta ($q\sim  1/r$; \eq\nr{Vr}; 
viewed from this
side $1/\epsilon$ is an ultraviolet divergence) 
of the spatial gluons~$A_i$ cancel against each other
in \eq\nr{fullVr}. 

\begin{figure}[t]


\centerline{%
\epsfysize=8.0cm\epsfbox{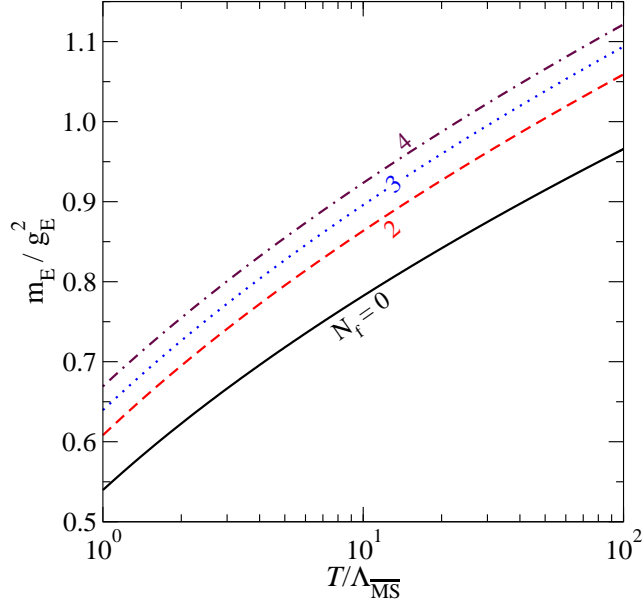}%
}


\caption[a]{\small
The parameter $\mE/\gE^2 = y^{1/2}$ from \eq\nr{y} for $\Nc=3$. 
}

\la{fig:rho}
\end{figure}

Carrying out suitable rescalings, \eq\nr{Seq} can be solved up to one 
transcendental number. We thus obtain
\be
 M_2 \approx 2 \mE + \frac{\gE^2\Nc}{2\pi}
 \Bigl( 
   0.60372466 - \fr12\!\ln\rho 
 \Bigr)
 \;, \la{M2_result}
\ee
where 
\be
 \rho \equiv \frac{\gE^2\Nc}{2\pi\mE} = \frac{\Nc}{2\pi y^{1/2}}
 \;. \la{rho}
\ee
At next-to-leading order in massless QCD
the ratio $y\equiv\mE^2/\gE^4$ is 
renormalization group invariant~\cite{adjoint},
and can be written compactly as 
\ba
 y \!\!&\approx&\!\! \frac{(2\Nc + \Nf)(11\Nc - 2\Nf)}{144\pi^2}
 \biggl[
  \ln\frac{4\pi T}{\Lambdamsbar} 
     - \gammaE + \frac{4\Nf\ln 2 - \Nc}{11\Nc - 2 \Nf}
     + \frac{5 \Nc^2 + \Nf^2 + 9 \Nf/2\Nc}{(2\Nc + \Nf)(11\Nc - 2 \Nf)}  
 \biggr] 
 \;. \nonumber \\[2mm] \la{y}
\ea
The corresponding $\mE/\gE^2 = y^{1/2}$ 
is plotted in \fig\ref{fig:rho} for $\Nc=3$.


%
\section{Summary and conclusions}

Comparing \eq\nr{M2_result} with \eq\nr{M2_def}, we obtain 
\be
 c_2 \approx 1.9467141
\ee
for $\Nc = 3$. Given that 
$\mE \ge 0.5 \gE^2$ (cf.\ \fig\ref{fig:rho}), 
the latter term in \eq\nr{M2_def} is always subdominant. 
This is in stark contrast to \eq\nr{M1_def}, in which the latter 
term, containing the non-perturbative coefficient $c_1 \approx 6.9$,  
dominates in the whole temperature range of phenomenological interest. 
Note that because $c_2 \ll c_3 \approx 10.0$,  
$M_2$ is in general also below $M_3$ 
(cf.\ \eq\nr{M3_def}) in the temperature range of \fig\ref{fig:rho}.
All three masses are plotted in \fig\ref{fig:masses}, both 
in units of $\gE^2$ and in units of $T$.

\begin{figure}[t]


\centerline{%
\epsfysize=7.0cm\epsfbox{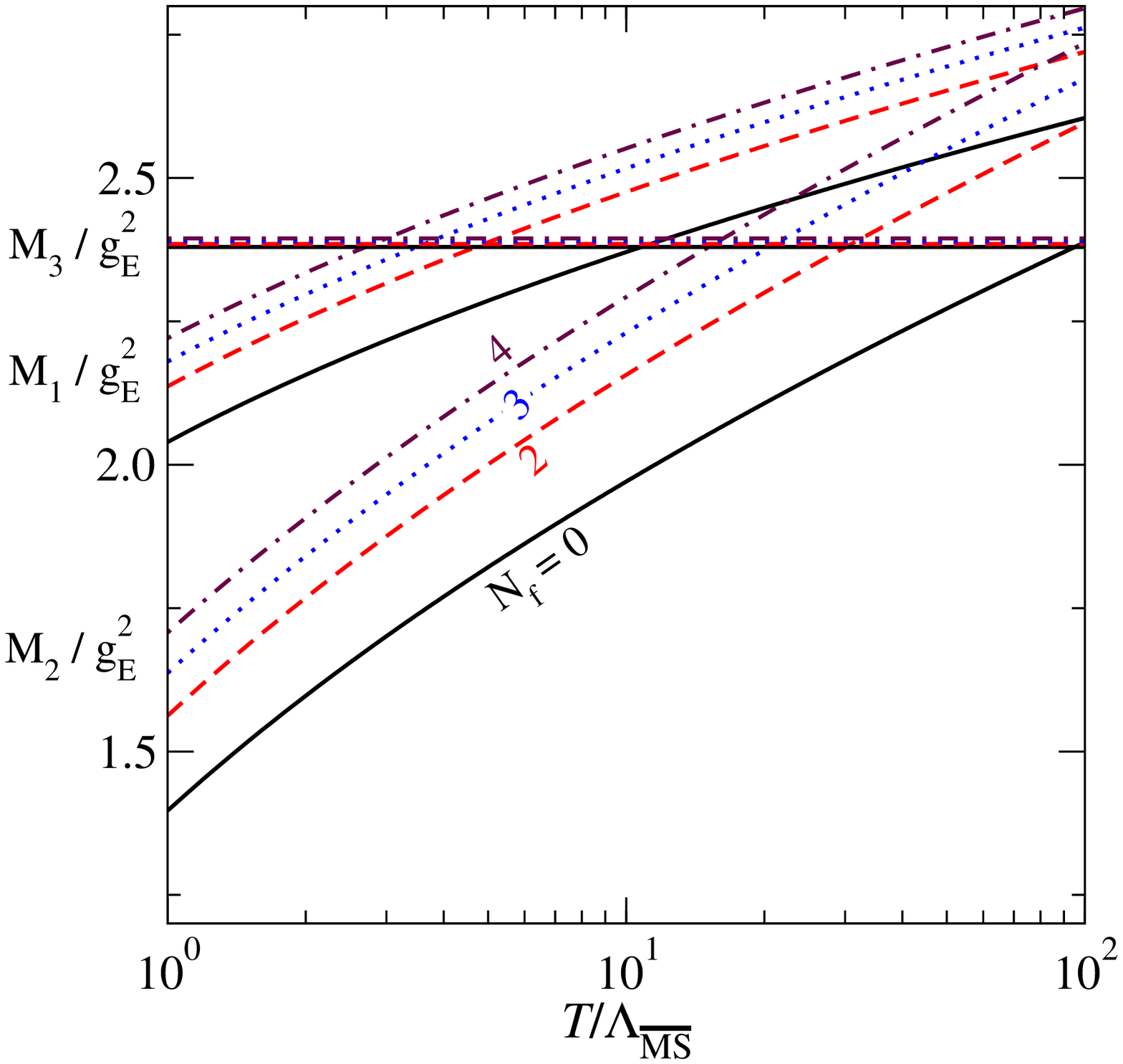}%
~~~\epsfysize=7.0cm\epsfbox{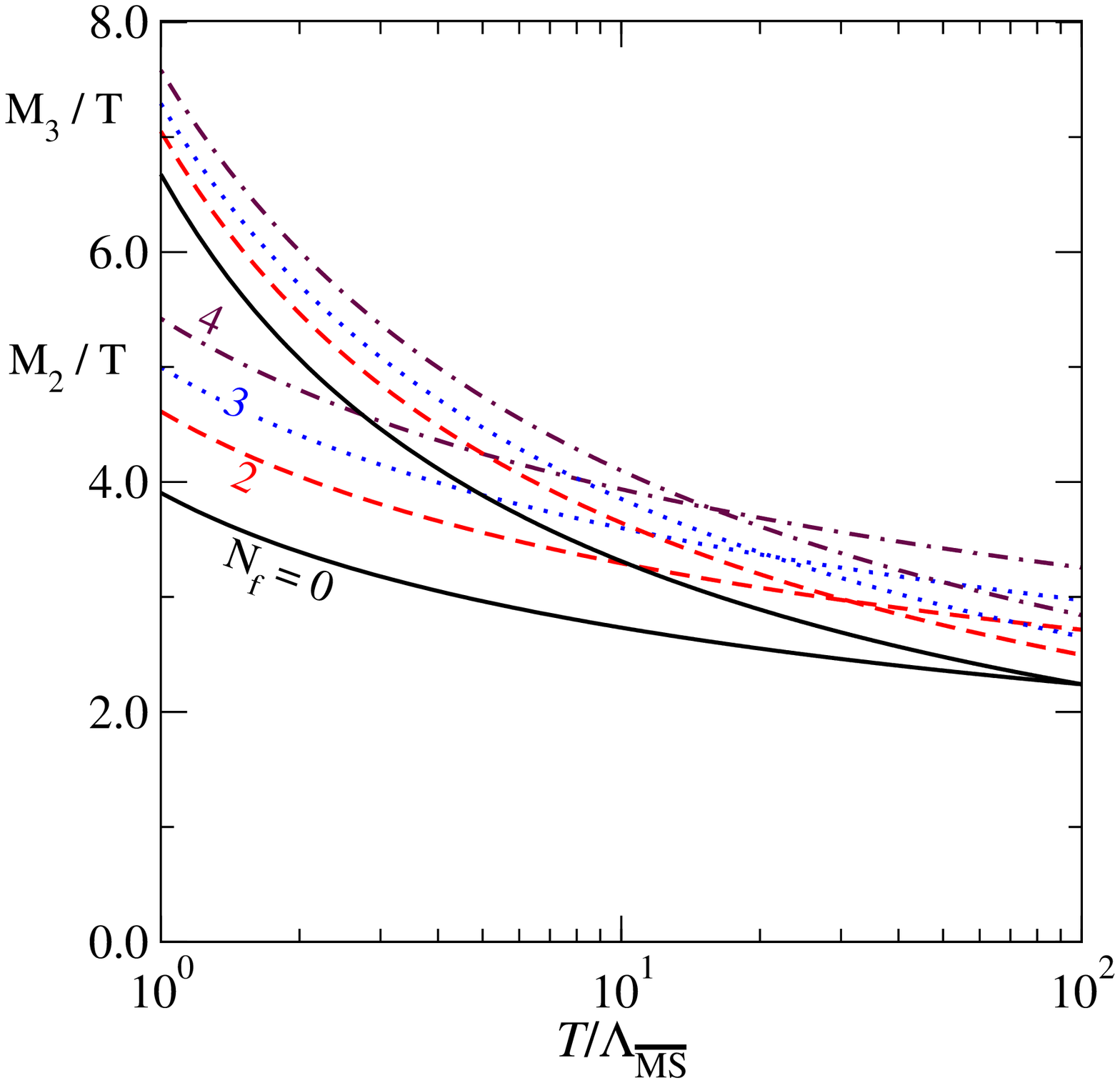}%
}


\caption[a]{\small
Left: The masses $M_1,M_2,M_3$ (cf.\ \eqs\nr{M1_def}--\nr{M3_def}), 
in units of $\gE^2$, for $\Nc=3$, after insertion of $\mE/\gE^2$ from
\eq\nr{y}. For better visibility, the four curves
for $M_3$, which are degenerate, have been slightly displaced from each other.
Right: The masses $M_2,M_3$, which belong to the same quantum number channel, 
after the insertion of the 2-loop value for $\gE^2 / T$ from ref.~\cite{gE2}. 
}

\la{fig:masses}
\end{figure}

To summarize, it appears 
understandable that $M_2$ represents the smallest screening
mass at realistic temperatures, 
because of the small perturbative coefficient $c_2 \approx 1.9$
in its next-to-leading order term, 
even though in the extreme
limit $\mE \gg \gE^2/\pi$ 
it eventually overtakes both $M_1$ and $M_3$, 
because a higher multiple of $\mE$'s appears in the leading term.
This observation, 
together with the explicit results in \fig\ref{fig:masses}, 
constitute the main points of this note.

We would like to stress, finally, that although our result for $M_2$ is
not meant to be quantitatively accurate at low temperatures, 
it nevertheless reveals an interesting pattern.   
For example, for $\Nf = 0$, \fig\ref{fig:masses} suggests
$M_2/T \approx 3 ... 4$ in the phenomenologically interesting
temperature range, while lattice measurements indicate
values  $M_2/T \approx 2.5 ... 3$~\cite{4d,hm2}, 
i.e.\ in the same ballpark but 
deviating downwards on the quantitative level. It appears, though, 
that this difference could at least partly be understood 
through higher order corrections {\em within} the EQCD effective
theory defined
by \eq\nr{eqcd}: for 
$T/\Lambdamsbar \approx 2$, the non-perturbative lattice 
measurements in ref.~\cite{mu} yielded 
$M_2/\gE^2 \approx 1.0, 1.3, 1.5, 1.6$ for
$\Nf = 0, 2, 3, 4$, respectively, 
and 
$M_1/\gE^2 \approx 1.7, 2.0, 2.1, 2.1$ for 
the same cases; these values lie consistently somewhat
below the perturbative estimates in
the left panel of \fig\ref{fig:masses}, 
resulting in a better accord with 4d lattice data.
Moreover, a similar overshooting of 
the $\rmO(\gE^2)$-corrected screening masses has been 
found for mesonic observables~\cite{lv,mv}.
So, it might be the general case that higher-order 
corrections, mostly from within three-dimensional EQCD dynamics, 
sum up to a negative correction to the next-to-leading order
expression for screening masses. 
This would imply that screening masses are different
in character from some dynamical quantities like the heavy quark  
momentum diffusion coefficient, in which case higher order corrections
appear to {\em add up} on top of the already large next-to-leading
order correction~\cite{mink2}.


%
\section{An open issue}

We end by briefly pointing out an open problem to which we 
have no solution. Consider the screening mass extracted
in four dimensions from the imaginary
part of the Polyakov loop; in EQCD this corresponds to $M(\tr[A_0^3])$. 
In weak coupling, 
$
 M(\tr[A_0^3]) = 3 \mE + \frac{\gE^2\Nc}{4\pi}
 (c_4 \ln\frac{\mE}{\gE^2} + c_5)
$. 
Even though this is heavier than $M_1$, 
particularly for $\Nf > 0$~\cite{mu}, 
it can be easily measured on the lattice~\cite{eh}, since the corresponding
operators have different quantum numbers. 
Therefore, it would be nice
to know the coefficients $c_4, c_5$. 
Though this is certainly a well-defined problem (for $\Nc > 2$), 
we have no clear idea about
how it could be solved in a systematic way. 
  (Probably the system can still be described by a non-relativistic 
   many-body Schr\"odinger equation with a certain three-body potential
   in it, but in the absence of an effective theory framework or an 
   explicit power-counting argument, it is difficult to know for sure 
   how to proceed without ambiguities.)
If the pattern found in this note 
continues, however, we might expect $c_4,c_5$ to be 
coefficients at most of order unity, such that 
the leading-order term would dominate even more
than in $M_2$.

%
\section*{Acknowledgements}

We are grateful to K.~Kajantie for discussions and 
to the BMBF for financial support under project
{\em Hot Nuclear Matter from Heavy Ion Collisions 
     and its Understanding from QCD}.
M.L.\ thanks the 
Institute for Nuclear Theory at the University
of Washington for its hospitality and the 
Department of Energy for partial support
during the completion of this work. 
M.V.\ was supported by the Academy of Finland, 
contract no.\ 128792. 


\appendix
\renewcommand{\thesection}{Appendix~\Alph{section}}
\renewcommand{\thesubsection}{\Alph{section}.\arabic{subsection}}
\renewcommand{\theequation}{\Alph{section}.\arabic{equation}}


\end{document}